# Ferroelectric Nematic Liquid Crystals as Charge Boosters for Triboelectric Nanogenerators

Jia-Yao Ye[1], Susanta Chakraborty[1], Karthick Subramani[1], Xing-Zhou Tang[1], Yan-Nan Xie[2,3,*] and Bing-Xiang Li[1,2,*]

[1]College of Electronic and Optical Engineering & College of Flexible Electronics (Future Technology), Nanjing University of Posts and Telecommunications, Nanjing 210023, China.

[2]State Key Laboratory of Flexible Electronics (LoFE), Nanjing University of Posts & Telecommunications, Nanjing 210023, China.

[3]Institute of Advanced Materials (IAM), Nanjing University of Posts & Telecommunications, Nanjing 210023, China.

* Corresponding authors: iamynxie@njupt.edu.cn; bxli@njupt.edu.cn

**Abstract:**

Driven by growing demand for clean energy, triboelectric nanogenerators (TENGs) have emerged as promising self-powered systems, yet achieving high charge density remains a critical challenge. In polymer dielectrics, triboelectricity can be further amplified by incorporating high-dielectric and polar materials for functional adaptability. Conventional dielectrics, including liquid crystals (LCs), offer limited improvement for triboelectrification, whereas the breakthrough ferroelectric nematic liquid crystals ($N_F$-LCs), with giant spontaneous polarization and high dielectric constant, act as highly effective charge boosters. Here, we introduce $N_F$-LC (DIO) as a functional filler in a PVDF-based TENG. Increasing DIO content progressively grows the electroactive phase and effective polarization in PVDF and defines a marked improvement in TENG's electrical performances through favourable dipolar alignment and strengthened charge-trap effects. The optimized composite film achieves an impressive open-circuit voltage of ~1.1 kV, short-circuit current of ~50 μA, and power density of 110 W/m² - seven times higher than pure PVDF. The device exhibits excellent charge-storage capability, powers over 500 LEDs without power management. This work establishes $N_F$-LC-based TENGs as a new platform for high-performance self-powered energy harvesting, linking soft matter physics with applied energy technology.

## 1. Introduction

With the inevitable exhaustion of non-renewable sources in the contemporary energy paradigm, an urgent need is to harness energy from clean and green sources. The critical conversion of micro-scale ambient energy sources into usable power has captivated both research and industry, serving as the key approach for unlocking the next generation of autonomous and self-powered systems. Intending to replace conventional batteries as an energy storage device, self-powered nanogenerators (NGs) have emerged as a transformative approach for self-governing operation of electronic devices, converting micro-mechanical environments[1-5]. Compared to several energy-harvesting technologies such as piezoelectric[6], triboelectric[7-12], electromagnetic[13] and flexoelectric nanogenerators[14-15], the triboelectric nanogenerators (TENGs) have attracted widespread attention due to their structural simplicity, flexibility, compatibility with diverse materials and high-power performance[7-12]. TENG utilizes the coupling of the frictional force between the surfaces of two materials with the electrostatic induction, in which a potential difference is generated between electrodes and triggers to transfer of the opposite triboelectric charges[16]. From the first pioneering discovery of TENG[17], enormous investigation has been accomplished on triboelectricity using conventional ceramic nanostructures such as ZnO, $BaTiO_3$, and lead zirconate titanate (PZT). However, recent studies have involved replacing the toxic lead-based materials with inorganic triboelectric materials like MXene, graphene, etc.[18-20], and also different polymeric systems, named as PDMS, PVA, PVDF, PI and TPU[21-25]. Among others, PVDF and its copolymers demonstrate the excellent potential/contribution for TENG owing to offering remarkable flexibility, light weight, and especially charge storing capability and stability of electrical performance over a long period. Despite a considerable improvement in material engineering, current challenges include improving the triboelectric output performance of TENG. Since the charge transferring or surface charge density is the key factor that influences the performance of TENG, researchers have been increasingly exploring techniques for boosting charge density to obtain effective efficiency. It has been well studied and demonstrated that inclusion of different nano-fillers[26-27], development of new functional materials with excellent performance[28], microstructure engineering on dielectric materials[29] and charge excitation[30-32] are affordably highlighted in TENGs for efficient energy management strategy. Specifically, the boosting of charge density improves the capacitance of TENG, which is governed by the mutual contribution of charge generation and dissipation. The former is mainly determined by the molecular interaction of the dielectric material, while the latter is influenced by the material's trap state and conductivity[33]. Existing research clarifies that controlling charge trapping and detrapping of dielectric materials greatly affects the induced triboelectricity[34]. Additionally, the inclusion of dielectric nano-fillers also endows the dipolar alignment, improves the dielectric permittivity and charge storing capability and thereby enhances the output performances[35-36]. Still, several limitations, including insufficient charge storage capability, excessive charge loss, and electronegativity differences between tribo-pairs, restrict the ability to meet the needs of ultrahigh performance and diverse functionality of dielectric TENGs. Therefore, the parameters for selecting dielectric materials like giant permittivity, polarization, dielectric coupling and proper electron cloud overlapping between tribo-pairs are essentially required to optimize for further boosting the charge trapping capability of PVDF-based tribo-negative material.

In contrast, the liquid crystals (LC) possessing intrinsic long-range anisotropic structural order, tunable polarizability, permittivity, reconfigurable anisotropic optical and electrical properties and soft-matter characteristics have emerged as a promising candidate for various types of ferroelectrics, electro-optic devices, including energy harvesting devices. Previous studies exhibit noticeable piezoelectric effects in columnar and smectic liquid crystals[37-39]. Moreover, different liquid crystalline polyarylates (LCPs) have been found to effectively improve the electrical performance of TENGs[28, 40-41]. However, most conventional LC systems exhibit weak dielectric permittivity and polar order, which limits the output performance of LC-based energy devices. The synthesis and stabilization of liquid crystals with strong polarity have therefore been a longstanding challenge. A breakthrough and widespread attention in soft matter was achieved with the discovery of ferroelectric nematic liquid crystals ($N_F$-LC), which exhibit fascinating properties such as huge spontaneous polarization at room temperature, giant dielectric constants, and sub-millisecond electro-optic switching, unlike conventional nematics[42-47]. The unique molecular architecture of these compounds enables robust unidirectional alignment and long-range polar order, providing a new route to exploit liquid crystal materials for high-efficiency energy harvesting. Previous studies already demonstrated the existence of a large piezoelectric coefficient in these $N_F$-LCs[39, 48], which is the resultant coupling effect of spontaneous electric polarization with long-range orientational order. However, their potential in inducing triboelectricity remains largely unexplored, offering a fertile area of interest. It is anticipated that such $N_F$-LC material with a large dielectric constant and high ferroelectric polarization should obviously boost the charge density mechanism of TENG.

Here, we employed the ferroelectric nematic liquid crystal, DIO, with PVDF to construct a combined ferroelectric-polymer triboelectric system, which serves as a tribo-negative layer in the contact-separation-mode of TENG. The incorporation of DIO promotes the β-phase formation, increases permittivity and remanent polarization, and strengthens interfacial dipolar coupling, which reflects a progressive enhancement nature of open-circuit voltage, short circuit current and power density of fabricated TENGs with an increase of DIO concentration. The results of fabricated TENG show that for the highest concentration, the open circuit voltage is about 1.1 kV, the short-circuit current is about 50 μA and the power density is 110 W/m$^2$ at an external force of 30 N, representing an order-of-magnitude improvement. The device exhibits self-enhanced and stabilized triboelectricity with high durability under mechanical cycling due to effective charge-trapping and interfacial polarization. The charge-storing capability of the fabricated high-performance TENG is also investigated and found to power up 500 white and blue LEDs without any external circuit management. Therefore, by leveraging the unique features of $N_F$-LCs, this work facilitates bridging the gap between fundamental soft-matter science and applied energy technology.

## 2. Results and Discussions

### 2.1. Experimental design and working principle

To prepare the $N_F$-LC-based TENG, firstly, the free-standing films of pure polymeric material PVDF and composites with four different weight concentrations of $N_F$-LC, DIO (3 wt%-named as FNLC-3 and 5 wt%-named as FNLC-5) are prepared via a DMF-based solvent drop-casting method without any mechanical and electrical treatment. A detailed discussion is provided in the experimental section. Each film was integrated into a vertical contact-separation-mode triboelectric nanogenerator (TENG), which is schematically illustrated in Fig. 1. Here, the pure PVDF and $N_F$-LC-doped PVDF films serve as the tribo-negative material. For the generation of triboelectrification, aluminum (Al) is chosen as the electro-positive material as well as the electrode, as shown in Fig. 1a,b. A conductive fabric is adhered over the tribo-negative PVDF film as the other electrode. In addition, a soft adhesive foam tape is purposely affixed to the back side of the two electrodes to fix them to the top and bottom acrylic plates. This entire structure induces triboelectrification during periodic mechanical movement of acrylic plates with a maximum separation of 5 mm. Since Al is electro-positive, it has a tendency to lose electrons, while the tribo-negative layer is prone to gain electrons. Initially, no electric potential is generated between the two triboelectric layers until they come into contact with each other [Fig. 1c(i)]. Once the PVDF or composite films come into contact with Al under external force, triboelectrification causes a transfer of surface charges between them to a maximum level [Fig. 1c(ii)]. After the contact, the surface of PVDF would have negative charges and Al should have positive charges depending on their electron affinities. During the separation [Fig. 1c(iii)], the induction of positive charges accumulates at the back electrode of PVDF or composite films. Effectively, the electrons flow from the bottom electrode to Al through the external circuit, followed by a positive voltage pulse till to equilibrium limit of charge transferring [Fig. 1c(iv)]. Again, when Al approaches the film [Fig. 1c(v)], electrons flow from Al to the bottom electrode, exhibiting a negative voltage pulse, balancing the potential difference. Cyclic operation of these steps processes the periodic positive and negative voltage pulses of the triboelectric signal.

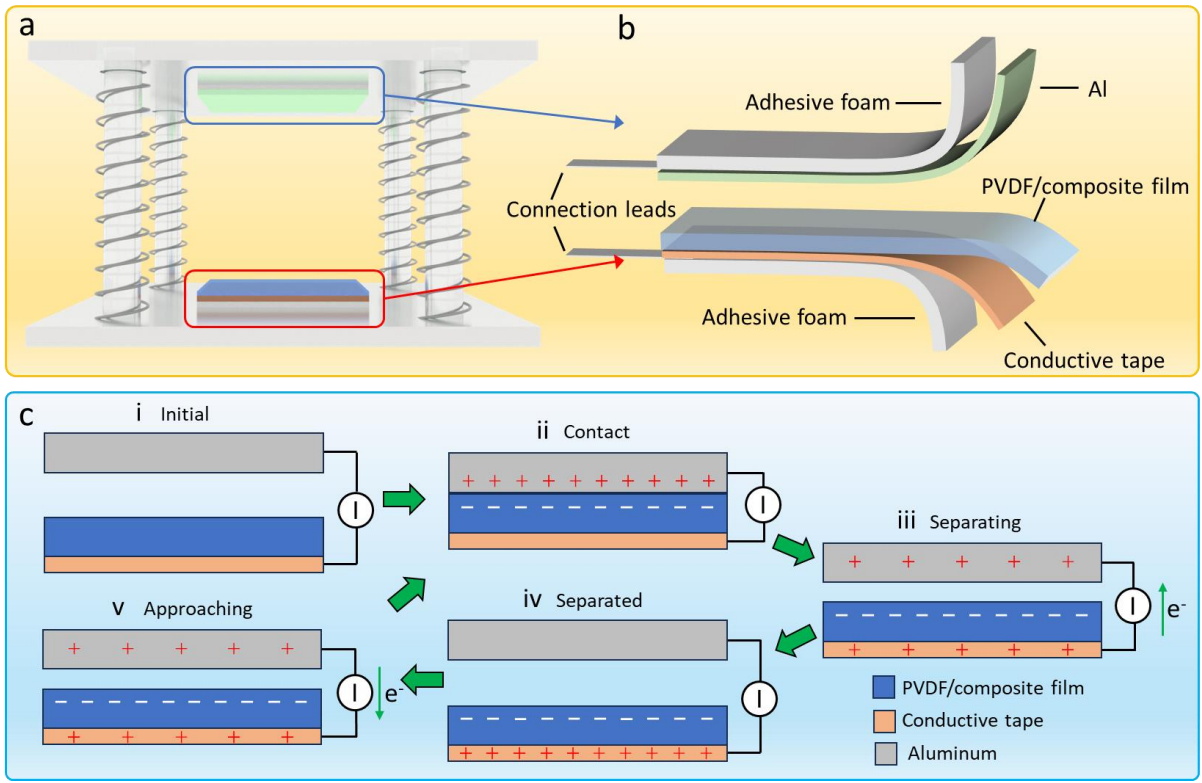

**Fig. 1.** (a) Actual structure of the experimental setup assembled with $N_F$-LC-based TENG, (b) Schematic illustration for the fabrication of the triboelectric pairs consisting of PVDF/composite film as tribo-negative layer and aluminum (Al) as tribo-positive layer, (C) Representation of working mechanism for charge generation and transfer during mechanical movement of Al layer.

## 2.2. Electrical performance and charge dynamics

The triboelectric performance was characterized using a mechanical linear motor operating between 12 and 45 Hz frequency under a 28 N impact force and a 4 mm gap between triboelectric pairs. Fig. 2a shows the open circuit voltage ($V_{OC}$) of PVDF film as well as composite films FNLC-2, FNLC-3 and FNLC-5, at a frequency of 40 Hz. It is evident that, under the same experimental conditions, PVDF film exhibits a maximum $V_{OC}$ of about 257 V, while FNLC-2, FNLC-3 and FNLC-5 generate 348 V, 630 V and 940 V, respectively. Therefore, the performance of the triboelectrification increases with an increase in the concentration of strongly polar $N_F$-LC material, indicating an enhancement of the surface charge density. Moreover, the frequency dependence of the performance for FNLC-5 film (Fig. 2b) shows that $V_{OC}$ increases progressively from 550 V to 1.1 kV with an increase in frequency from 12 Hz to 45 Hz, with no observable saturation. Similar behavior is also found for other composite films. The increase in frequency defines a shorter time for charge transferring, which effectively favors a large number of electron flows triggered by the electrification effect of the electronegative FNLC-5 film. The obtained short-circuit current ($I_{SC}$) measured across 10 MΩ resistance, shown in Fig. 2c, increases from 19.8 µA to a maximum of 49.2 µA with increasing frequency from 12 Hz to 45 Hz. Results for other composite films represent the identical behavior. Following Ohm's law, the output voltage should be proportional to external resistance, while the current is inversely proportional to resistance. The obtained variation of output voltage ($V_{SC}$) and short-circuit current ($I_{SC}$) for a series of connecting load resistances (inset of Fig. 2c) elucidates a similar nature, although there is a little variation below the 1 MΩ resistance. Furthermore, we determined the power density for all the films under different connected load resistances at a frequency of 40 Hz and depicted in Fig. 2d. It is clearly seen that the obtained power density possesses a maximum value at around 10 MΩ resistance for all the samples. When the resistance of the external circuit was very small or large, the triboelectric charges tended to be consumed inside the TENG film with time, leading to lower charge transfer to the electrodes. Remarkably, with increasing the concentration of $N_F$-LC, the maximum power density rises from 16.7 W/m² for pure PVDF to 29.7 W/m², 53.5 W/m² and 110.7 W/m² for FNLC-2, FNLC-3 and FNLC-5 films, respectively, which are significantly higher than the majority of triboelectric materials. These findings demonstrate that the addition of $N_F$-LC enhanced the induction of electric dipoles suitably aligned for better triboelectric synergetic performance.

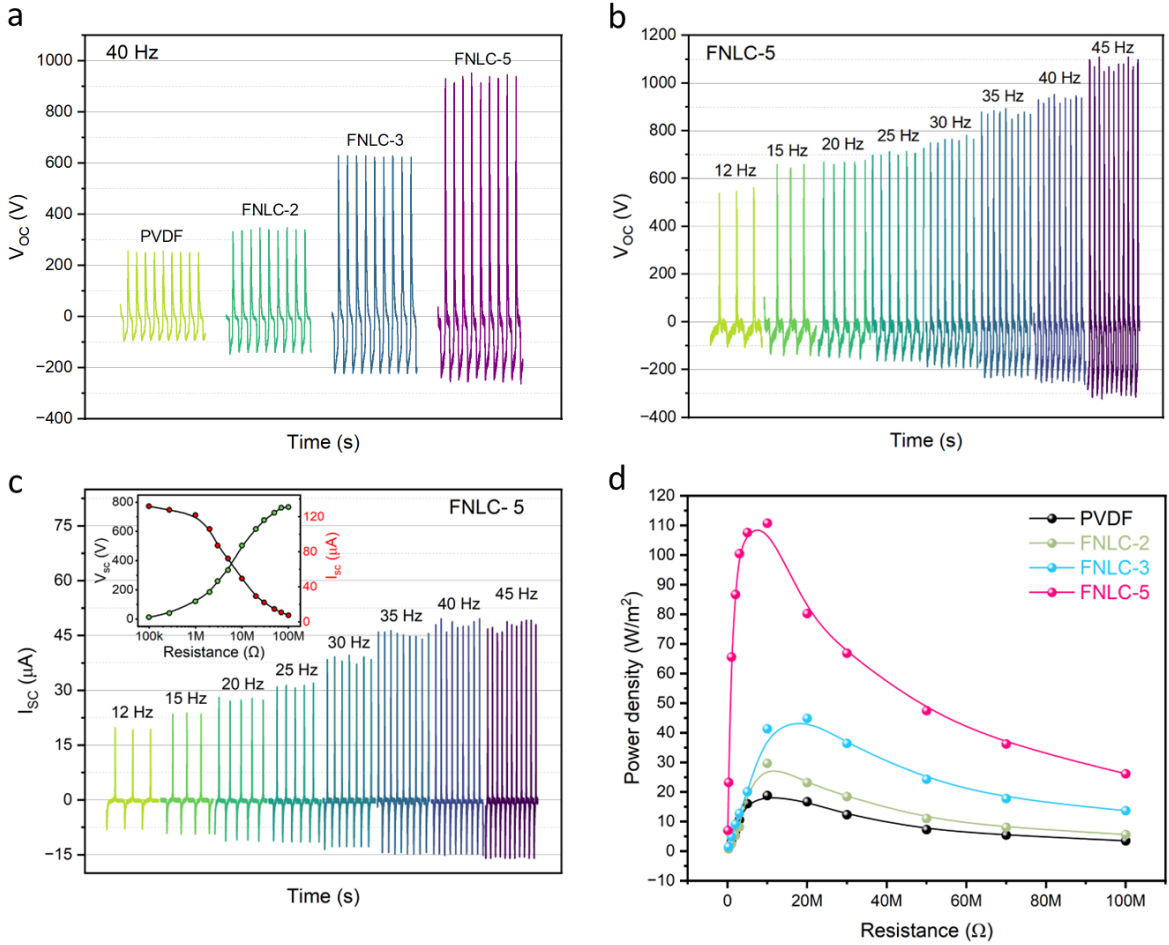

**Fig. 2.** The electrical characteristics of the developed $N_F$-LC-based TENG. (a) Variation of open circuit voltage ($V_{OC}$) for pure PVDF film and composite films at frequency 40 Hz, (b) Dependence of $V_{OC}$ for film FNLC-5 on frequency of operation. (c) Dependence of short-circuit current ($I_{SC}$) for film FNLC-5 on frequency of operation. The inset shows the obtained values of short-circuit voltage and current at different resistances. (d) Variation of power density with load resistance for all the fabricated TENG films.

The output performance of $N_F$-LC-based TENG is primarily governed by the obtained surface charge density. Analysis of output current during the continuous operation of TENG, the current gradually increases from a lower value to a saturated equilibrium, a phenomenon attributed to the progressive accumulation and trapping of surface charges within the PVDF matrix. The intrinsic charges immediately after contact of the triboelectric pair are small and gradually increase to a saturation level. Therefore, the charges are deeply trapped within the dielectric material, which is quite evident for the PVDF matrix[34,49]. Specifically, the induced charges during contact electrification transport through the material, are trapped by the shallow traps and some of them migrate to the deep traps (need higher energy to escape). Effectively, a large number of charges are accumulated inside until sufficient energy is gained to escape from the deep trap state[34,50]. The saturation of charge accumulation occurs by balancing charge leakage, trapping, detrapping and transportation processes, followed by a stable output from the TENG. This charge accumulation saturation time decreases with increasing frequency, leading to a fast injection of charges. However, compared to the $N_F$-LC-based films with the pure PVDF film, the saturation time increases with an increase in concentration. This suggests that the charge trapping sites are quite large in the $N_F$-LC-based films compared to pure PVDF, proving the influence of the molecular interaction of $N_F$-LC. In addition to charge trapping, the dipole-dipole interaction and dielectric hysteresis facilitate the self-polarization phenomenon inside the material, which also increases the surface charge density of the TENG film[30, 51-53]. By doping the $N_F$-LC material with PVDF, a more polarized state emerges, resulting in an enhanced charge trapping capability of the PVDF film during contact electrification. Therefore, the effective coupling of dielectric polarization and charge accumulation significantly increases the number of triboelectric charges in the triboelectric material during TENG operation, generating a large surface charge density until a saturation level is reached.

## 2.3. Operational stability

In order to determine the effect of the applied force between triboelectric pairs on the electrical performance of $N_F$-LC-TENG, measurements were carried out at different forces. Fig. 3a defines the variation of output current at different applied forces from 0.5 N to 35 N. It is observed that the maximum short-circuit current increases gradually from 5.7 µA to 61 µA with an increase in applied force from 0.5 N to 35 N. This result elucidates that upon an increase in force, it effectively increases the contact area between tribo-pairs, which in turn influences a larger number of charges to be induced in the system, contributing to a higher electrical output performance of the $N_F$-LC-based TENG. Furthermore, to test the stability and durability of the prepared $N_F$-LC-TENG, the output current has been measured for the FNLC-5 film under 40 Hz frequency and 30 N applied force. Fig. 3b defines that $I_{SC}$ initially increases to a saturation level and maintains stability over a long time. However, the maximum current experiences a certain fluctuation or drop of ~3.2% for a continuous operation of 16500 cycles. This small change was observed due to the suppression of foam placed at both the backside of $N_F$-LC-PVDF film and the Al film, for operating at high pressure and frequency. Notably, the same film again responds to the almost maximum current limit even after 1 month. These results reveal that without any degradation, the $N_F$-LC-TENG film exhibits excellent stability and durability for practical applications.

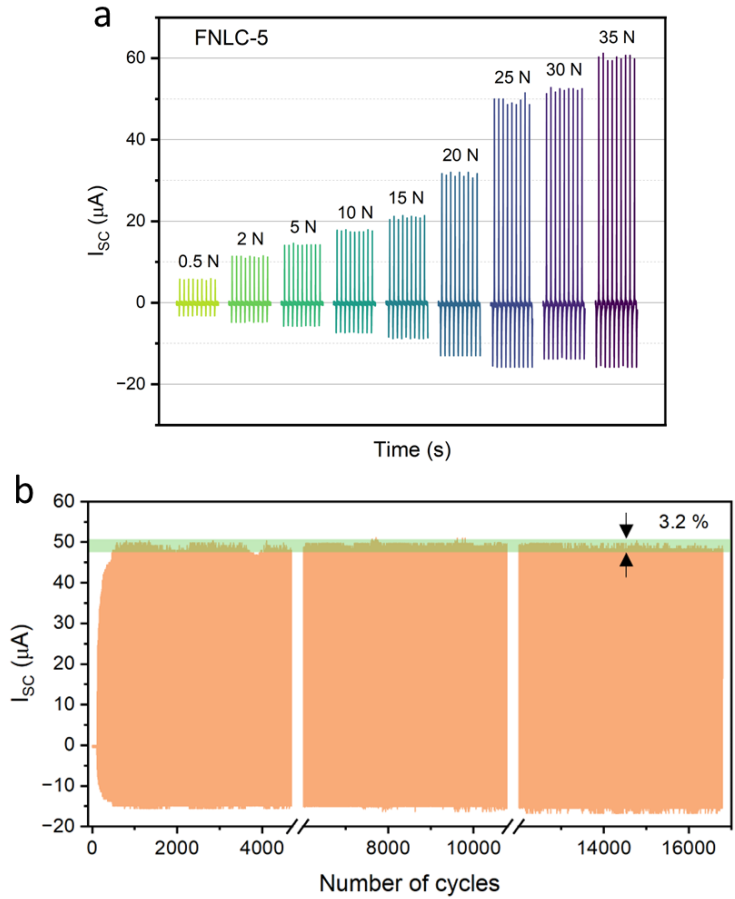

**Fig. 3.** (a) Variation of short-circuit current ($I_{SC}$) for FNLC-5 with different applied forces at 40 Hz frequency. (b) Stability and durability test of short-circuit current for FNLC-5 over 16500 cycles under 30 N force and 4 mm separation.

### 2.4. Functional demonstration

From the perspective of the electrical characteristics, the addition of $N_F$-LC filler collectively improved the performance of the fabricated TENG. Among all the samples, the FNLC-5 film demonstrates efficient voltage along with power characteristics, which can be utilized for further application devices. To realize as an active power source, first, the output of the TENG (with FNLC-5 film) is rectified by a single diode (half-wave rectifier) and then connected with LED lights in a series connection. Fig. 4a shows the associated circuit diagram and Fig. 4b illustrates the real-time photograph of powering LED lights. It was observed that the device assembled with FNLC-5 film, working at 40 Hz with a force of 28 N, can drive ~500 white and blue LED lights connected in series without any power management. The stable brightness under periodic actuation indicates uniform charge transfer with negligible internal loss. To assess energy-storage performance, the rectified output from the bridge rectifier was directed to a charging circuit (green dotted box in Figure 4c) containing a FR207 diode, 15 mH inductor, and 150 pF input capacitor ($C_{in}$). The FNLC-5 TENG operating at 40 Hz charged 10 µF, 22 µF, 47 µF and 100 µF capacitors up to 49.5 V, 27.3 V, 14.5 V and 5.28 V, respectively, within 120 seconds. However, the smaller 3.3 µF capacitor reached 50 V in just 49 seconds (Figure 4c and Figure 4d). The rapid charging and high terminal voltage highlight the efficient conversion of mechanical motion into stored electrical energy.

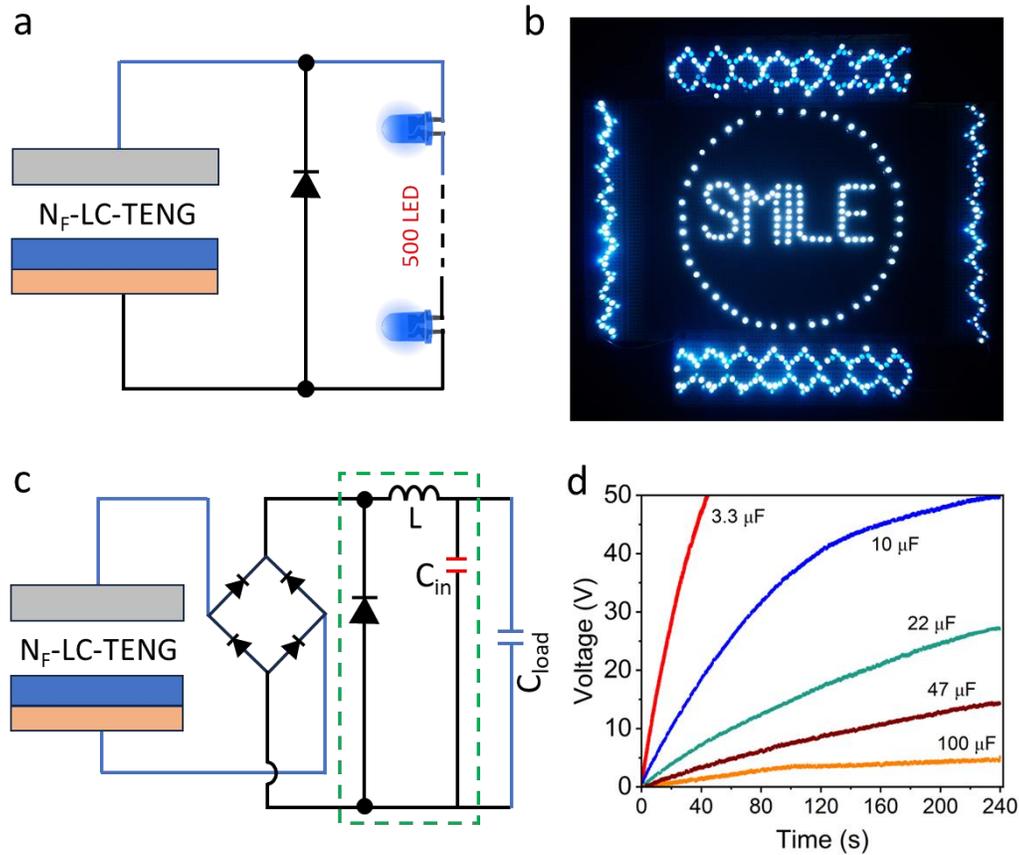

**Fig. 4.** (a) The circuit diagram to power the LEDs using prepared $N_F$-LC-TENG, (b) Real-time photograph of 500 LEDs powered by FNLC-5 film, (c) The circuit diagram to experimentally charge different capacitors, (d) The variation of charged voltage with time for different capacitors by $N_F$-LC-TENG operated at 40 Hz.

Compared with conventional piezoelectric or electromagnetic generators, the $N_F$-LC-based TENG exhibits higher power density under low-frequency stimuli (< 50 Hz) owing to cooperative polarization and efficient charge trapping. These results confirm that the combined $N_F$-LC-PVDF composites transform molecular-scale ferroelectric order into macroscopic electromechanical conversion, offering a practical pathway toward flexible, self-powered, and multifunctional electronic systems.

## 3. Conclusion

This work focuses on improving the tribo-electrification of the contact-separation mode TENG by incorporating the highly polar ferroelectric nematic liquid crystal ($N_F$-LC) with PVDF polymer as the negative triboelectric layer. The fabricated concentration-dependent $N_F$-LC-doped TENG exhibited significantly higher triboelectric characteristics in terms of current, power density compared to pure PVDF film, without compromising stability and flexibility. Due to the intriguing properties of $N_F$-LC, it is apparent that a finite growth of the electroactive phase occurs upon doping with PVDF polymer, along with the enlargement of dielectric and polarization properties in composite films, due to feasible dipolar orientation and effective surface charge polarization. Significantly, the mutual interaction of polar groups increases the net dipole moment, higher band gap energy level, as well as charge trap parameters, which in turn boost the electrical output performance of TENG. The composite film doped with 5 wt% exhibits the open circuit voltage of 1.1 kV, short-circuit current of 50 µA and the power density of 110 W/m$^2$ at a force of 30 N, representing approximately one order of magnitude larger with respect to pure PVDF film. Results showed that the fabricated device possesses a stabilized electrical output over more than 16000 mechanical cycles and can drive 500 LEDs directly without any external circuit design. Beyond its immediate energy-harvesting performance, this work demonstrates how long-range molecular polarization in soft-matter systems can be harnessed to amplify macroscopic electrical output. The molecular design principle presented here offers a versatile platform for developing next-generation flexible, self-powered devices and provides a link between ferroelectric liquid-crystal physics and functional energy technologies.

## 4. Experimental Section

**Materials:** The ferroelectric nematic liquid crystal ($N_F$-LC), DIO was procured from Nanjing Shuxin Technology Co., Ltd. PVDF (polyvinylidene fluoride) powder was purchased from Bide Pharmatech Ltd, Shanghai and utilized as a primary component for the preparation of composite films. The solvent DMF (N, N-dimethylformamide) was purchased from Shanghai Macklin Biochemical Technology Co., Ltd. All the chemicals were used without any further purification. All other materials not explicitly mentioned were obtained from local sources.

**Preparation of sample and TENG Film:** PVDF powder of the amount 0.5 g was dissolved in 10 ml DMF and magnetically stirred for 1 hour at a temperature of 50 °C. The homogeneous solution appears dense and transparent, which was kept in a vacuum chamber for 2 hours for degassing. The solution was drop-casted onto a glass plate and kept in the vacuum oven at a temperature of 78 °C for 3 hours. After cooling down to room temperature, the PVDF film was peeled off by dipping the glass plate into deionized water. Finally, this film was dried in the oven at 100 °C. The preparation process of $N_F$-LC-doped PVDF films adhered to the same steps; however, to obtain a uniform mixture, the desired amount of $N_F$-LC material was mixed with PVDF-DMF solution and stirred for 2 hours at a temperature of 60 °C. Free-standing films of pristine PVDF and composites (of thickness ~40-60 µm) with four different weight concentrations of DIO (2 wt%-named as FNLC-2, 3 wt%-named as FNLC-3 and 5 wt%-named as FNLC-5) are prepared without any mechanical and electrical treatment.

**Fabrication of TENG device:** To prepare the TENG device, the fabricated films were first cut into pieces with dimensions of 1.5 cm × 1.5 cm. The conductive fabric of the same dimension was pasted over the PVDF/$N_F$-LC-doped PVDF films and copper wires were connected as leads. For the positive triboelectric layer, an aluminum sheet of dimension 1.5 cm × 1.5 cm (thickness ~30 µm) is used and a connection wire was attached to this aluminum sheet. To operate in contact and separation mode of these triboelectric pairs, a structural design was accomplished by using acrylic sheets and rods as well as mechanical springs (Figure 1). The as-prepared tribo-positive and negative pairs were pasted at the center of the flat surface of two rectangular acrylic sheets (5 cm × 5 cm) with the help of double-sided foam tape. This foam allows for restricting excessive impact on the triboelectric layers and also avoids unnecessary noise. The acrylic sheets were kept separated by four mechanical springs at four corners, supported by solid acrylic columns. This structural design delivers the smooth contact and separation mode of triboelectric pairs within a maximum separation range of 2 cm. However, this separation was kept fixed at 4 mm during the study, unless mentioned. Further, we controlled the frequency of the contact-separation operation within a range of 12 Hz to 45 Hz by using a linear motor attached to the entire structural design of the TENG device.

**Characterization and Measurement:** To assess the triboelectric performance, a lab-made spring-modulated platform (Fig. 1a) was prepared using acrylic sheets and a mechanical linear motor was used to attain the frequency range between 12 and 45 Hz. The electrical characteristics were measured by a digital oscilloscope, Rigol DS1202Z-E. The applied force on the TENG device was measured and controlled with the help of a pre-calibrated commercial force sensor (FSR 402) operated by Arduino and PC.

# Acknowledgements

The work is supported by the National Key Research and Development Program of China (No. 2022YFA1405000, No. 2021YFA1202904), the Natural Science Foundation of China (No. 62375141, No. 62288102), the Natural Science Foundation of Jiangsu Province, Major Project (No. BK20243067), the Basic Research Program of Jiangsu Province (No. BK20250039, BK20243057), the Agricultural Science & Technology Independent Innovation Fund of Jiangsu Province (No. CX(23)3124).